\documentclass[epj,nopacs]{svjour}

\usepackage{amsmath,amssymb,graphicx,bm}
\usepackage{color}
\usepackage{multicol}        % used for the two-column index
\usepackage[bottom]{footmisc}% places footnotes at page bottom

\newcommand{\beq}{\begin{equation}}
\newcommand{\eeq}{\end{equation}}
\newcommand{\bea}{\begin{eqnarray}}
\newcommand{\eea}{\end{eqnarray}}
\newcommand{\ba}{\begin{align}}
\newcommand{\ea}{\end{align}}
\newcommand{\bfig}{\begin{figure}}
\newcommand{\efig}{\end{figure}}

\newcommand{\D}{\displaystyle}

\newcommand{\gev}{\, \text{GeV}}

\begin{document}

%\begin{frontmatter}

%
\title{Constraining the
low energy pion electromagnetic form factor with space-like data}
\author{B.\ Ananthanarayan \and S. Ramanan}
%\email{anant@cts.iisc.ernet.in}
%\email{suna@cts.iisc.ernet.in}

\institute{Centre for High Energy Physics,
Indian Institute of Science, Bangalore\ 560 012, India.\\ {\email{anant@cts.iisc.ernet.in} \\ \email{suna@cts.iisc.ernet.in}}}        

\date{\today}
%
%%%%%%%%%%%%%%%%%%%%%%%%%%%%%%%%%%%%%%%%%%%%%%%%%%%%%%%%%%%%%%%%%%%%%%%%%%%%%%%%%%%%%%%%%%%%%%%
\abstract{The pionic contribution to the $g-2$ of the muon involves a certain 
integral over the the 
modulus squared of $F_\pi(t)$, the pion electromagnetic form 
factor.  We extend techniques that use cut-plane analyticity properties of 
$F_\pi(t)$ in order to account for present day estimates of the pionic 
contribution and experimental information at a finite number of 
points in the space-like region. Using data from several experiments over 
a large kinematic range for $|t|$, we find bounds on the 
expansion coefficients of $F_\pi(t)$, sub-leading to the 
charge radius. 
The value of one of these coefficients in chiral perturbation 
theory respects these bounds. Furthermore, we present a sensitivity 
analysis to the inputs.  A brief comparison with results in the literature 
that use observables other than the $g-2$ and timelike data is presented.}

\maketitle

%\end{frontmatter}
%%%%%%%%%%%%%%%%%%%%%%%%%%%%%%%%%%%%%%%%%%%%%%%%%%%%%%%%%%%%%%%%%%%%%%%%%%%%%%%%%%%%%%%%%%%%%%%%%
\section{Introduction}
\label{introduction}
General principles such as analyticity and unitarity have been used
to derive useful constraints on form factors (for a review see 
ref.~\cite{SinghRaina}).  In particular, useful lower bounds were 
obtained on the pionic contribution
to the $(g-2)$ of the muon (muon anomaly).  For recent
reviews on the current status of this quantity, see 
refs.~\cite{hep-ph/0703049,hep-ph/0703125}.
Consider the following low-energy expansion for the pion electromagnetic
form factor $F_\pi(t)$,
\beq
F_{\pi}(t) = F_\pi(0) + \D\frac{r_\pi^2}{6} \,t + c \,t^2 + d \,t^3 + \cdots,
\label{fpi.eqn1}
\eeq  
where $r_\pi$ is the pion charge radius, $c$ and $d$ are Taylor coefficients, with units of $\gev^{-4}$ and $\gev^{-6}$ respectively. For a recent discussion, refer~\cite{hep-ph/0203049}.  
From general principles, discussed in the next section,
and using the normalization $F_\pi(0)=1$, immediately gives a 
lower bound of $\approx 1.6 \times 10^{-9}$. Imposing the experimental value of
$r_\pi$ further improves the bound to $2.9\times 10^{-9}$~\cite{PHRVA.D4.1558}.
Caprini~\cite{Caprini}
has more recently shown that if one were to assume that the
pionic contribution is less than $I = 75\times 10^{-9}$, then these principles 
yield constraints on the higher expansion coefficients, $c,\, d$
which translate into an allowed ellipse in the $c-d$ plane. These results, as well as the results we present here, hold as long as the true pionic contribution is less than the 
value of $I$ we have assumed. 
If the pionic contribution is indeed significantly smaller, then the true 
allowed region would be a proper subset of the region we isolate.  

The inclusion of experimental data from the time-like
region, the phase, the modulus, and the phase as well as the modulus
each improves the bound on the muon anomaly~\cite{SinghRaina}. 
Using $I$ as an input, Caprini 
has used dispersive techniques and time-like
data, and has considered 
other physical observables such as certain QCD polarization functions,
in addition to the muon anomaly, to constrain further the allowed
regions in the $c-d$ plane~\footnote{Applications to form factors
in semi-leptonic decays of such general methods in a modern context
have also been recently considered~\cite{CapriniBourrely}.}.
We note that
the parameter $c$ itself has been determined to 2-loop accuracy in 
chiral perturbation theory~\cite{hep-ph/0203049} and has a value of 
$4.49\, \gev^{-4}$, which is safely accommodated in the
allowed ellipse of Caprini~\cite{Caprini}.  

An important source of experimental information are the values
of the electromagnetic form factor in the space-like region. These
have been reported in electroproduction experiments conducted in 
the 1970's (see refs.~\cite{Bebek1},~\cite{Brown},~\cite{Bebek2}). 
Raina and Singh~\cite{RainaSingh} used this information
to produce improved bounds on the muon anomaly (see also~\cite{NenRes}). New experimental data is now available (see refs.~\cite{Dally1},~\cite{Dally2},~\cite{Amendolia} and~\cite{nucl-ex/0607007}). In this work, we will make use of  some of the data sets above.

Our main
purpose in this work is to demonstrate that the framework of Raina and Singh
can be effectively used to obtain constraints in  
the $c-d$ plane, using  the pionic contribution to the muon anomaly as an input. This is an algebraic framework, which is clear cut 
and transparent. It provides an important consistency check on the allowed
regions isolated by Caprini.  

In sec.~\ref{formalism} we briefly review the dispersive formalism
and describe the implementation of
space-like constraints and the method by which we isolate  
the region in the $c-d$ plane. Here, we provide the theoretical
framework and also a discussion on the present experimental status,
as they are both required for this purpose. In sec.~\ref{results} we  
present our results using data 
from~\cite{Bebek1},~\cite{Brown},~\cite{Bebek2} and more recent data from~\cite{Amendolia} and~\cite{nucl-ex/0607007}. 
We present a detailed discussion of our results and our conclusions in sec.~\ref{conclusions}.

%%%%%%%%%%%%%%%%%%%%%%%%%%%%%%%%%%%%%%%%%%%%%%%%%%%%%%%%%%%%%%%%%%%%%%%%%%%%%%%%%%%%%%%%%%%%%
\section{Formalism}
\label{formalism}

We recall the formalism presented in ref.~\cite{RainaSingh}.
The pion contribution to the muon anomaly is given by:
\beq
a_{\mu} (\pi^{+} \pi^{-}) = \D\frac{1}{\pi} \int_{t_{\pi}}^{\infty} dt\, 
\rho(t) |F_{\pi}(t)|^2
\label{amu.eqn1}
\eeq
where $t_\pi = 4 m_{\pi}^2$ is the branch point of the pion form factor and
\beq
\rho(t) = \D\frac{\alpha^2 m_{\mu}^2}{12 \pi} \frac{(t - t_\pi)^{3/2}}{t^{7/2}} K(t) \ge 0
\label{rho.eqn1}
\eeq
where,
\beq
K(t) = \int_0^1 du\, (1-u) u^2 (1- u + \D\frac{m_\mu^2 u^2}{t})^{-1}.
\label{rho.eqn2}
\eeq
Using the following map from the $t$-plane, which is cut from $t_\pi$ along the real $t$ axis, to the complex $z$-plane (region $|z|<1$),
\beq
\D \frac{z-1}{z+1} = i \sqrt{\frac{t - t_\pi}{t_\pi}},
\label{ztmap.eqn1}
\eeq 
and the definitions:
\beq
f(z) = F_{\pi}(t)
\label{defn.eqn1}
\eeq
\beq
p(z) = \rho(t)
\label{defn.eqn2}
\eeq
the pionic contribution to the muon anomaly can be written as
\beq
a_{\mu} (\pi^{+} \pi^{-}) = \D\frac{1}{2 \pi} \int_0^{2 \pi} d\theta \,w(\theta) |f(e^{i \theta})|^2
\label{amu.eqn2}
\eeq
where,
\beq
w(\theta) = 4 m_\pi^2 \sec^2 (\theta/2) \tan (\theta/2)\ p\,(e^{i\theta}) \ge 0. 
\label{amu.eqn3}	 
\eeq	
We now consider a function $h(z)$ defined as:
\beq
 h(z) = f(z) w_\pi(z)
\label{hz.eqn1} 
\eeq
where,
\beq 
w_\pi(z) = \exp\left[\D\frac{1}{4 \pi} \int_0^{2 \pi} d\theta\, \D\frac{e^{i\theta}+z}{e^{i\theta}-z} \ln w(\theta)\right].
\label{hz.eqn2}
\eeq
%Equivalently, we can write eqn.~(\ref{hz.eqn2}) as,
%\beq
%w_\pi(z) = \exp\left[\D\frac{1}{2\pi} \int_0^{\pi} d\theta\, \D\frac{1 - z^2}{1 + z^2 - 2z \cos(\theta)} \ln|w(\theta)| \right]
%\label{hx.eqn}
%\eeq
%which is real for real $z$.
Then eqn.~(\ref{amu.eqn2}) can be written as:
\beq
a_{\mu}(\pi^{+} \pi^{-}) = \D\frac{1}{2 \pi} \int_0^{2 \pi} d\theta \,|h(e^{i \theta})|^2	
\label{amu.eqn4}
\eeq
Now $h(z)$ is analytic within the unit circle $|z| < 1$ and for real $z$, $h(z)$ is real. Therefore $h(z)$ can be expanded as follows:
\beq
h(z) = a_0 + a_1 z + a_2 z^2 + \cdots
\label{hz.eqn3}
\eeq
where $a_0, a_1 \cdots$ are real coefficients.
Therefore, in the analytic region, $|z| \le 1$, $a_{\mu}(\pi^{+} \pi^{-})$ can be written as:
\beq
a_{\mu} (\pi^{+} \pi^{-}) = a_0^2 + a_1^2 + \cdots
\label{amu.eqn5}
\eeq
The expansion coefficients $a_n$ can be obtained from a Taylor expansion of 
the function $h(z)$ in terms of $f(z)$ and $w_{\pi}(z)$. The coefficients are 
given by
\beq
a_0 = h(0) = w_\pi(0),
\eeq
\beq
a_1 = h^{\prime}(0)= w_\pi^{\prime}(0) +\D\frac{2}{3} r_\pi^2 t_\pi w_\pi(0),
\eeq
\bea
a_2 & = &\D\frac{h^{\prime \prime}(0)}{2!}=\frac{1}{2}\left[
	w_\pi(0)\left(-\frac{8}{3} r_\pi^2 t_\pi + 32 \,c \, t_\pi^2\right) \right] \nonumber \\ 
	 &+& \frac{1}{2} \left[2 w_\pi^{\prime}(0)\left(\frac{2}{3} r_\pi^2 t_\pi\right) +w_\pi^{\prime \prime}(0) \right], 
\eea
and
\bea
a_3 &=& \D\frac{h^{\prime \prime \prime}(0)}{3!} =
\D\frac{1}{6} \left[ w_\pi(0)\left(12 r_\pi^2 t_\pi - 384\, c\, t_\pi^2 + 384 \,d \, t_\pi^3\right) \right] \nonumber \\
&+& \D\frac{1}{6}\left[3 w_\pi^{\prime}(0)\left(-\frac{8}{3} r_\pi^2 t_\pi + 32 \,c \, t_\pi^2\right)\right] \nonumber \\
&+&\D\frac{1}{6} \left[
2 w_\pi^{\prime \prime}(0) r_\pi^2 t_\pi + w_\pi^{\prime \prime \prime}(0) \right].  
\eea
In our treatment, the expansion coefficients satisfy
\beq
\sum_{i = 0}^{\infty} a_i^2 = I
\label{cons.eqn2}
\eeq
Given $I$, eqn.~(\ref{cons.eqn2}) yields constraints on 
the expansion coefficients of
the form factor. 
Including up to the second (third) derivative 
for $F_{\pi}(t)$ results in constraints for $c$ ($c$ and $d$). 
It may be pointed out that Caprini's result on $c$ and $d$ shown in Fig. 1 (dashed lines) of
ref.~\cite{Caprini} is obtained precisely in this manner.

The constraints on the expansion coefficients of interest may be
significantly improved through the inclusion of experimental information
on the form factor.  Our objective in this work is to study the effect
of including experimental information coming from the space-like region.
In order to meet this objective, we first extend the formalism that
has been presented in ref.~\cite{RainaSingh}.  In that work, space-like
constraints were used to obtain lower bounds on the muon anomaly, denoted as $I_{{\rm min}}$.

We observe that $N$
space-like constraints are linear constraints that may be expressed as:
\beq
A_n = \sum_{i=0}^{\infty} a_i c_i^{(n)}
\label{space_cons.eqn1}
\eeq
where $n = 1, 2, \cdots N$. Such constraints can be implemented through the 
technique of Lagrange multipliers, by setting up the Lagrangian
\beq
L= \D\frac{1}{2} \sum_{i=0}^\infty
 a_i^2 +\sum_{n=1}^N \alpha_n (A_n - \sum_{i=0}^\infty a_i c_i^{(n)})
\label{space_cons.eqn2}
\eeq
where in eqn.~(\ref{space_cons.eqn2}), we consider only finite number of expansion coefficients.
The Lagrange equations yield:
\beq
a_i = \sum_{n=1}^N \alpha_n c_i^{(n)} 
\label{space_cons.eqn3}
\eeq
and 
\beq
I=\sum_{n=1}^N \alpha_n \sum_{i=0}^\infty a_i c_i^{(n)}.
\label{space_cons.eqn4}
\eeq
It then follows that 
\beq
I_{{\rm min}} = \sum_{n=1}^N \alpha_n A_n,
\eeq
where
\beq
A_n = \sum_{m=1}^N \alpha_m \sum_{i=0}^\infty c_i^{(m)} c_i^{(n)}.
\label{space_cons.eqn5}
\eeq
All the $\alpha_n$s may be eliminated to yield a determinantal equation
for $I$
\beq
\left|
\begin{array}{c c c c}
 I  &  A_1  &  A_2  & ...    \\
  A_1  & \sum c_i^{(1)} c_i^{(1)} & \sum c_i^{(1)} c_i^{(2)} & ... \\   
  A_2  & \sum c_i^{(2)} c_i^{(1)} & \sum c_i^{(2)} c_i^{(2)} & ... \\   
. & ... & ... & ... \\
. & ... & ... & ... \\
. & ... & ... & ... \\
\end{array}\right| =0.
\eeq
For the case at hand, where we wish to specify $a_0,\, a_1,\, a_2, a_3$
and the value at space-like points $h(x_i),\, i=1,2,3...$, where $x_i$ is real,
the determinantal equation reads,
\beq
\left|
\begin{array}{c c c c c c c c}
I & a_0 & a_1 & a_2 & a_3 & h(x_1) & h(x_2) & ...\\
a_0 & 1 & 0 & 0 & 0 & 1 & 1 & ...\\
a_1 & 0 & 1 & 0 & 0 & x_1 & x_2 & ... \\
a_2 & 0 & 0 & 1 & 0 & x_1^2 & x_2^2 & ...\\
a_3 & 0 & 0 & 0 & 1 & x_1^3 & x_2^3 & ... \\
h(x_1) & 1 & x_1 & x_1^2 & x_1^3 & (1-x_1^2)^{-1} & (1-x_1 x_2)^{-1} & ...\\
h(x_2) & 1 & x_2 & x_2^2 & x_2^3 & (1-x_2 x_1)^{-1} & (1-x_2^2)^{-1} & ...\\
. & ... & ... & ... & ... & ... & ... & \\
. & ... & ... & ... & ... & ... & ... & \\
. & ... & ... & ... & ... & ... & ... & \\
\end{array}\right|=0.
\label{det.eqn1}
\eeq
In the above merely retaining the first two rows and columns gives the first bound of Palmer, given in the Introduction, while retaining the first three rows and columns yields the second of Palmer's bounds.
Raina and Singh~\cite{RainaSingh} use the value of $F_\pi(0) = 1$ and $r_\pi$ to obtain the lower bound to $a_\mu(\pi^+\pi^-)$. This amounts to dropping out the rows and columns corresponding 
to $a_2$ and $a_3$ in the eqn.~(\ref{det.eqn1}) which are related to the expansion coefficients $c$ and $d$.
 Instead, providing an input to $I$ in eqn.~(\ref{det.eqn1}), gives us an allowed region in
the $c-d$ plane.
Dropping the row and column corresponding to $a_3$ would
result in determining an allowed region for $c$ alone, which we
pursue in the next section for purposes of illustration.

In the next subsection, we provide a discussion on the present day
experimental information that is utilized in our
study.  This information spans an impressive range of energies, {\it viz.}
$|t|$.
While in principle there is no limit
to the number of constraints, in practice the uncertainties 
in the experimental determination and sensitivity of the determinantal 
equation limits this number. The reason for such restrictions is the extreme sensitivity of the matrices to these experimental uncertainties
as their dimensions increase.  This sensitivity is particularly
severe for small values of $|t|$.  As a result, at such values
we are able to implement at most two constraints, while data
from higher energies allows us to implement up to three constraints.
However, the data from smaller values of $|t|$ provide
more stringent bounds for fixed number of 
constraints.
Thus we see a fairly complex interplay
between the energy regime that we can use and the number of constraints
we are able to implement.  
  
%%%%%%%%%%%%%%%%%%%%%%%%%%%%%%%%%%%%%%%%%%%%%%%%%%%%%%%%%%%%%%%%%%%%%%%%%%%%%%%%%%%%%%%%%%%%%%%

\subsection{Space-like data}
\label{experiments}

We will begin with the
data in the space-like region that was used in the work of Raina and Singh~\cite{RainaSingh}, that came from
measurements in the seventies. 
We shall refer to these as
the Brown data and Bebek data respectively.  The data we
use from these sets is given in Tables~\ref{table_bebek},~\ref{table_brown}. The tables also list the values of $z(t) = x(t)$, which is the map from the $t$ plane to the disc $|z| \le 1$ ($x(t)$ lies in the range $\left[-1 \le x(t) \le 0 \right]$), for the chosen data points and the corresponding value of $h(z) = h(x)$ (refer eqn.~(\ref{hz.eqn1})). To make our notations clear, we refer to the data points corresponding to a particular $|t|$ as $x_1$, $x_2$ and so on, in the ascending order of magnitude of $|t|$. The tables also show the experimental errors in the data. 

\begin{table*}
	\begin{center}
	\caption{Space-like data from Bebek et.al~\cite{Bebek2}}
	\label{table_bebek}
		\begin{tabular}{lllll}
		\hline\noalign{\smallskip}
		  & $t$($-Q^2$) [$\gev^2$] & $F_{\pi}(t)$ & $x(t)$ & $h(x) \times 10^{-5}$ \\
		\noalign{\smallskip}\hline\noalign{\smallskip}  	
		1 & -0.620 & 0.453 $\pm$ 0.014 & -0.499 & 3.057\\ 
		2 & -1.216 & 0.292 $\pm$ 0.026 & -0.606 & 2.035\\ 
		3 & -1.712 & 0.246 $\pm$ 0.017 & -0.655 & 1.716\\ 
		\noalign{\smallskip}\hline
	\end{tabular}  
	\end{center}
	\vspace*{0.2cm}
	\begin{center}
	\caption{Space-like data from Brown et.al~\cite{Brown}}
	\label{table_brown}
		\begin{tabular}{lllll}
		\hline\noalign{\smallskip}
		 & $t$($-Q^2$) [$\gev^2$] & $F_{\pi}(t)$ & $x(t)$ & $h(x) \times 10^{-5}$ \\
		\noalign{\smallskip}\hline\noalign{\smallskip}     	
		1 & -0.294 & 0.606 $\pm$ 0.028 & -0.372 & 3.775\\ 
		2 & -0.795 & 0.380 $\pm$ 0.013 & -0.540 & 2.608\\ 
		\noalign{\smallskip}\hline
	\end{tabular} 
	\end{center}
	\vspace*{0.2cm}
	\begin{center}
	\caption{Space-like data from Tadevosyan et al.~\cite{nucl-ex/0607007}}
	\label{table_jlab}
		\begin{tabular}{lllll}
		\hline\noalign{\smallskip}
		 & $t$($-Q^2$) [$\gev^2$] & $F_{\pi}(t)$ & $x(t)$ & $h(x) \times 10^{-5}$ \\
		\noalign{\smallskip}\hline\noalign{\smallskip}    	
		1 & -0.600 & 0.433 $\pm$ 0.017 & -0.494 & 2.915\\ 
		2 & -1.000 & 0.312 $\pm$ 0.016 & -0.576 & 2.163\\
		3 & -1.600 & 0.233 $\pm$ 0.014 & -0.645 & 1.626\\  
		\noalign{\smallskip}\hline
		\end{tabular}
	\end{center}
	\vspace*{0.2cm}
	\begin{center}
	\caption{Space-like data from Amendolia et.al~\cite{Amendolia}}
	\label{table_aman}
		\begin{tabular}{lllll}
		\hline\noalign{\smallskip}
		 & $t$($-Q^2$) [$\gev^2$] & $F_{\pi}(t)$ & $x(t)$ & $h(x) \times 10^{-5}$ \\
		\noalign{\smallskip}\hline\noalign{\smallskip}  	
		1 & -0.131 & 0.807 $\pm$ 0.015 & -0.242 & 4.454\\
		2 & -0.163 & 0.750 $\pm$ 0.016 & -0.275 & 4.286\\ 
		\noalign{\smallskip}\hline
		\end{tabular}
	\end{center}
\end{table*}

More recent data come from four sources.
The first of these is ref.~\cite{Dally1}  
which is from a Fermilab experiment
(F1).  Here the range is $0.03 {\gev}^2 \le -t \le 0.07 {\gev}^2$. 
Even at the high end, we are way below the lowest energy of Brown.  
The next set of
data come from another Fermi lab experiment (F2) (ref.~\cite{Dally2}),  
covering the  range of $0.037 {\gev}^2 \le -t \le 0.094 {\gev}^2$. 
The CERN NA7 experiment provides very accurate data~\cite{Amendolia}
in the  range $0.014 {\gev}^2 \le -t \le 0.26 {\gev}^2$ (by Amendolia et.al). 
There is an overlap region between this last set and
the Brown data and the two are consistent. The final data 
we use come from the JLab experiments (Tadevosyan et al., ref.~\cite{nucl-ex/0607007}).

In our study we choose to work with recent data from Amendolia et.al~\cite{Amendolia} that cover both the low and intermediate energy range (note that the data from~\cite{Dally1,Dally2} cover a smaller energy range; hence our choice) and the data from the JLab experiments (Tadevosyan et al., ref.~\cite{nucl-ex/0607007}) that cover a higher range of $|t|$. Using really small $|t|$ values lead to numerical instabilities as the entries in the determinant (eqn.~(\ref{det.eqn1})) become small. 

Tables~\ref{table_bebek},~\ref{table_brown},~\ref{table_jlab},~\ref{table_aman} show the data that is used in our analysis to constrain the expansion coefficients $c$ and $d$ of the form factor $F_{\pi}(t)$.
The data from various experiments are chosen so as to give reliable numerical results. We find data in lower $|t|$ to constrain the expansion coefficients better but we cannot include more than two space-like constraints; while the higher $|t|$ region gives a weaker bound, which can be improved by increasing the number of space-like constraints.

%%%%%%%%%%%%%%%%%%%%%%%%%%%%%%%%%%%%%%%%%%%%%%%%%%%%%%%%%%%%%%%%%%%%%%%%%%%%%%%%%%%%%%%%%%%%%%%%%%%%%%%%%%%%%%%%%%%
\section{Results}
\label{results}

In our results, we use $I = 75 \times 10^{-9}$ as an input, as quoted in~\cite{Caprini}. This value is certainly greater than a recent estimate for the hadronic contribution to $a_{\mu}^{\rm had} = 69.2 \times 10^{-9}$ given in~\cite{hep-ph/0703125}. Pionic contribution is expected to be around $70\%$ of the total hadronic value. Our value for $I$ gives a conservative upper bound to the pionic contribution to $a_\mu$. 

\begin{figure*}
\begin{center}
\includegraphics*[angle = 0, width = 0.84\textwidth, clip = true]{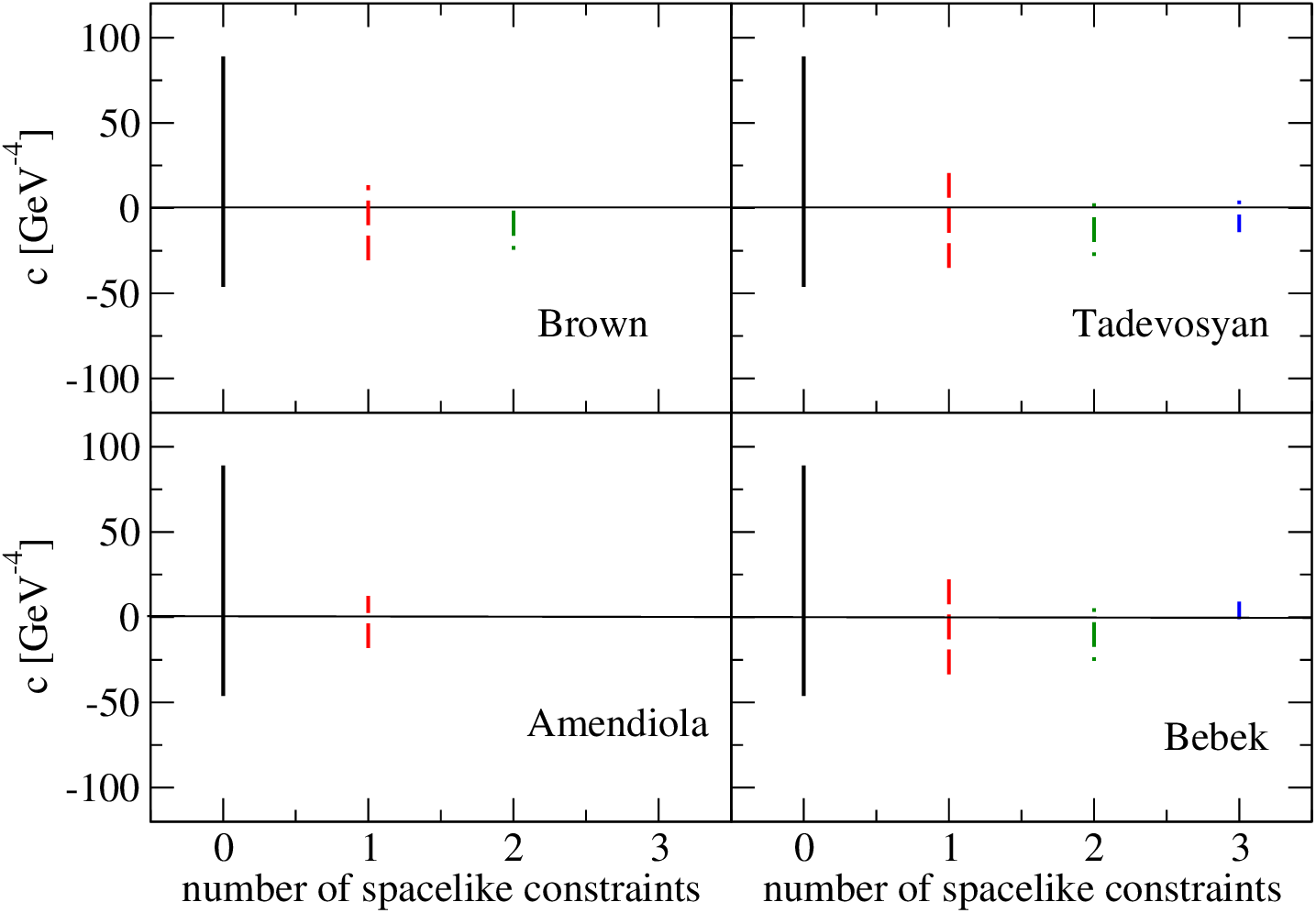}
%\vspace*{1cm}
\caption{Constraints on $c$ starting with just information on $F_{\pi}(0) = 1$ and $r_{\pi}^2$ and incorporating space-like constraints. Note that the allowed values of $c$ decrease as we include more space-like constraints.}
\label{cd_space1.fig1}
\end{center}
\end{figure*}

We begin by constraining only the expansion coefficient $c$ which amounts to dropping out the row and column corresponding to $a_3$ in eqn.~(\ref{det.eqn1}). We determine the bounds on $c$ in stages, starting  
with only the pion charge radius $r_{\pi}$ and the normalization of $F_\pi(t)$ at $t = 0$ (i.e., no space-like constraints) and then incorporate space-like constraints successively. The results are shown in fig.~(\ref{cd_space1.fig1}), where we have used data from tables~\ref{table_bebek},~\ref{table_brown},~\ref{table_jlab}, and~\ref{table_aman}. For any data set, we start with the smallest value of $|t|$ for one space-like constraint and include more constraints in increasing magnitude of $|t|$. All the data sets show the trend that inclusion of more space-like constraints 
in this manner, improve the bounds on $c$. The largest range for $c$ in units of $\gev^{-4}$ 
is obtained when no space-like constraints are used.  This agrees
with the result of Caprini (fig.(1) in~\cite{Caprini}).  Note however, that for data from larger
values of $|t|$, the bound
on $c$ is significantly weaker, compared to data from smaller values of $|t|$ for a fixed number of space-like constraints.

It is worth investigating the sensitivity of the bounds to the errors in the data. The Brown data shifts the bounds on $c$ to negative values when two space-like constraints are used. This shift is sensitive to the error bounds on $F_\pi(t)$. If we use as input $F_\pi(t_1) + 0.028$  (table~\ref{table_brown}) for the first constraint and $F_\pi(t_2) - 0.013$ for the second constraint, then the upper bound on $c$ is positive. Here, $t_1$ and $t_2$ represent the first and second data point respectively in table~\ref{table_brown}. We present further sensitivity analysis later in this section. Similarly for the data from Amendolia et.al~\cite{Amendolia}, we see that one space-like constraint can be incorporated using central values for $F_\pi(t)$; but varying this input over the error bounds given in table~\ref{table_aman}, we can use up to two constraints (not shown here). Unless othewise specified, all results presented in this section use the central values of the form factor data given in tables~\ref{table_bebek},~\ref{table_brown},~\ref{table_jlab}, and~\ref{table_aman}..

\begin{figure*}
\begin{center}
\includegraphics*[angle = 0, width = 0.84\textwidth, clip = true]{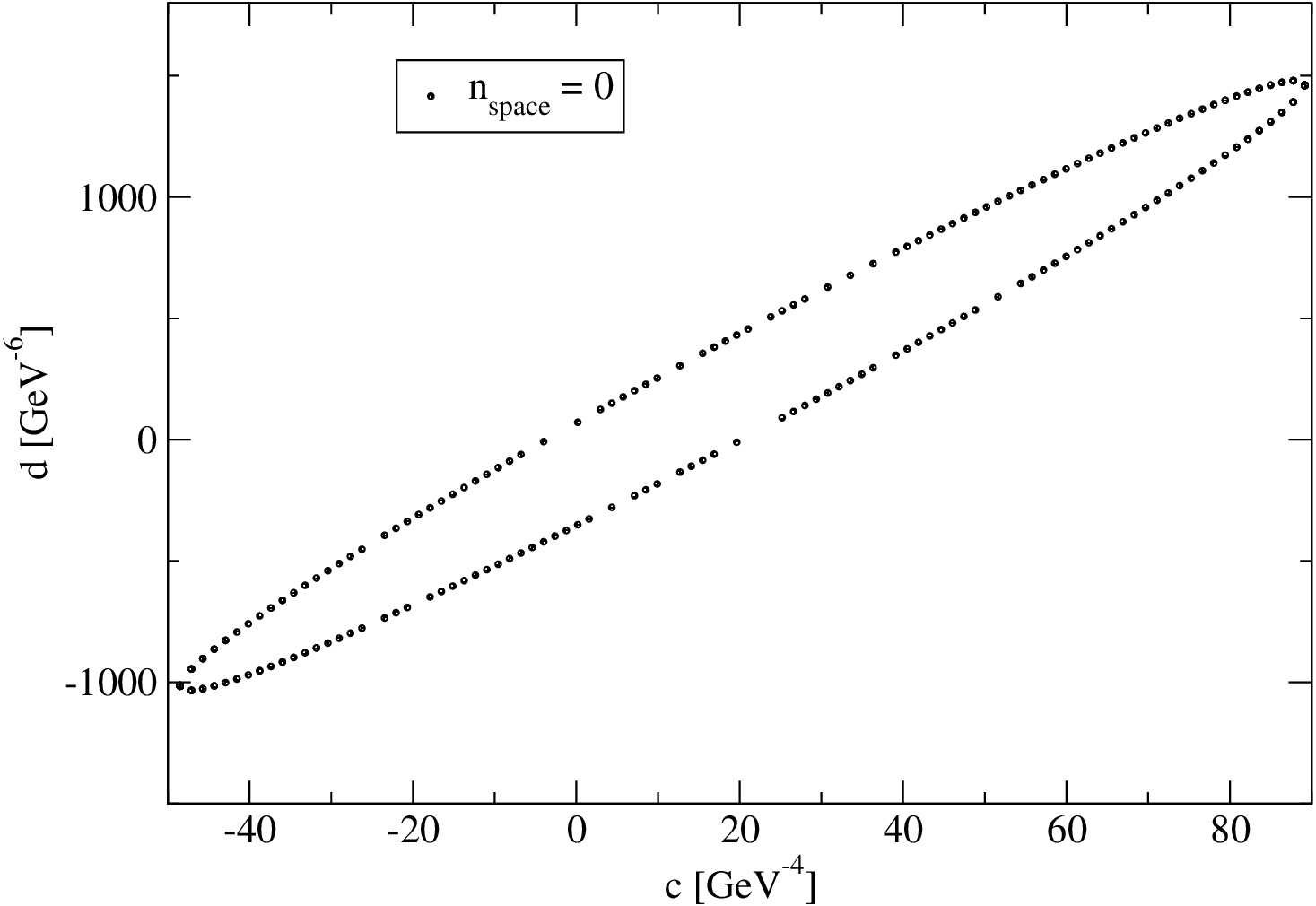}
\caption{Constraints on $c$ and $d$ using the normalization $F_\pi(0) = 1$ and the pion charge radius $r_\pi$.}
\label{cd_sep.fig0}
\end{center}
\end{figure*}

%Introducing
%constraints further restrict the range of $c$. In fig.~\ref{cd_space1.fig1} 
%we have used the data in tables~\ref{table_bebek},~\ref{table_brown},~\ref{table_jlab},~\ref{table_aman}.  
%We impose space-like constraint starting with the smallest value of $|t|$
%We start with the smallest value of $t$ for
%each data set when one space-like constraint is taken, and then the smallest
%and next to smallest values of $|t|$ when two are taken, and so on. 
%All the data sets show the trend that inclusion of more space-like constraints 
%in this manner, improve the bounds in $c$. 

\begin{figure*}
\begin{center}
\includegraphics*[angle = 0, width = 5.0in, clip = true]{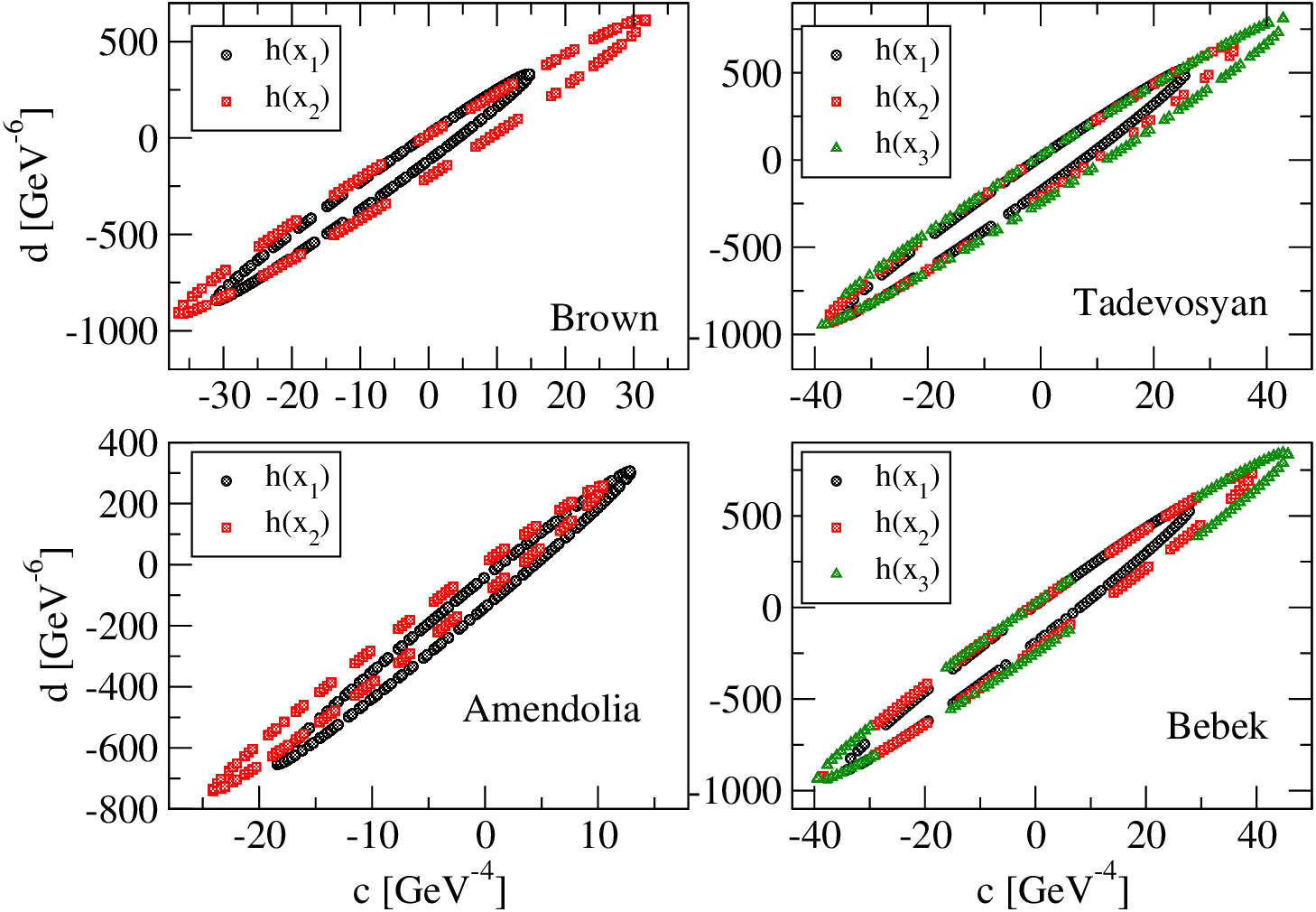}
\caption{Constraints on $c$ and $d$ using one space-like constraint. Here we start with the smallest $|t|$ value and the corresponding value of $h(z) = h(x_1)$ and study the variation of the bounds on $c$ and $d$ as we use data from higher the $|t|$ region.}
\label{cd_space1.fig2}
%\vspace*{0.1cm}
%\includegraphics*[angle = 0, width = 4.0in, clip = true]{pion_form_factor_coeffn_RS_N3_c_d_data_compare_final_nspace.eps}
%\caption{Constraints on $c$ and $d$ as the number of space-like constraints are increased.}
\end{center}
%\label{cd_nspace.fig3}
\end{figure*}

\begin{figure*}
\begin{center}
\includegraphics*[angle = 0, width = 0.84\textwidth, clip = true]{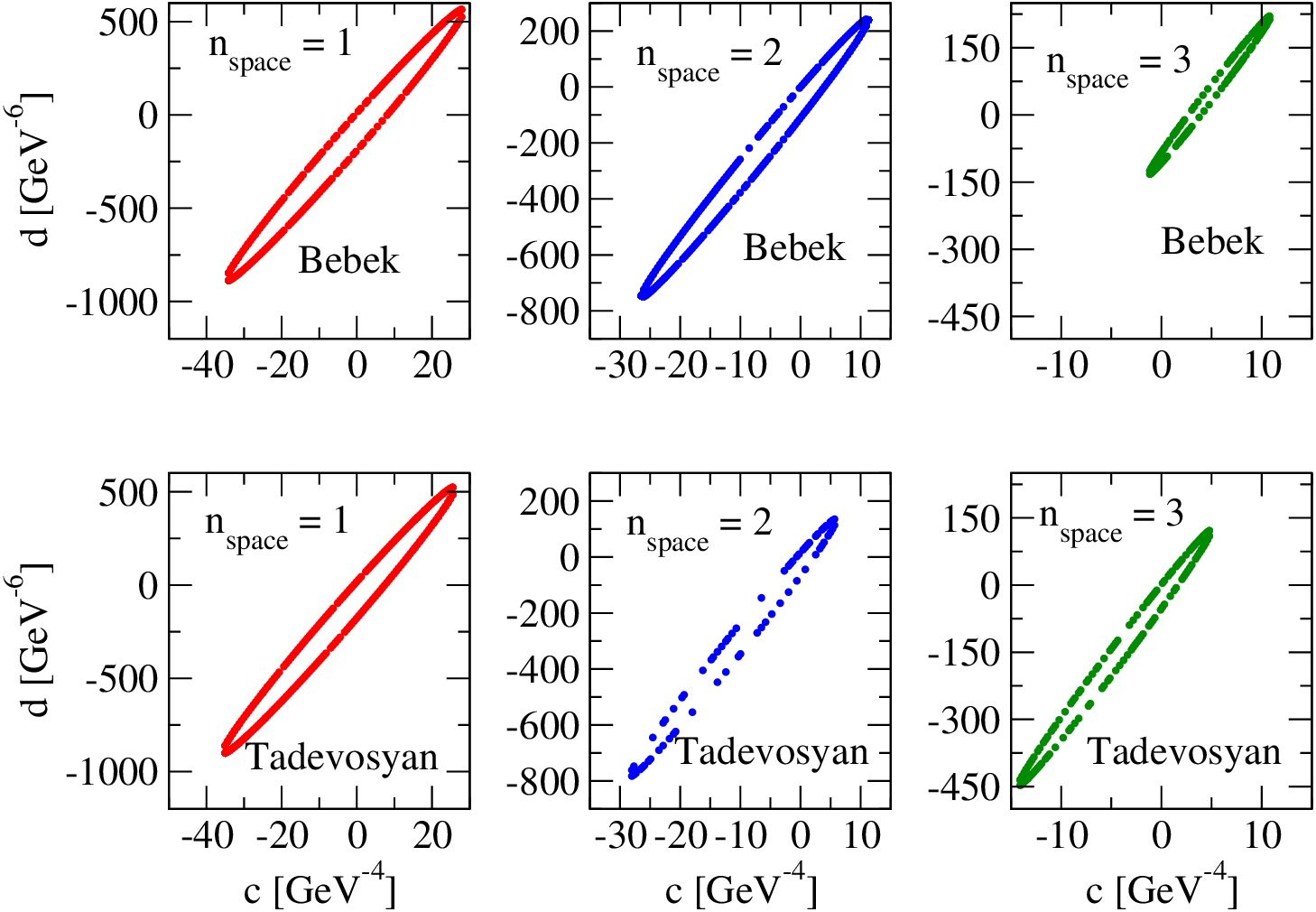}
\caption{Constraints on $c$ and $d$ as the number of space-like constraints are increased. Here the space-like data is taken from refs.~\cite{Bebek2},~\cite{nucl-ex/0607007}.}
\label{cd_sep.fig1}
\end{center}
\end{figure*}

\begin{figure*}
\begin{center}
\includegraphics*[angle = 0, width = 0.84\textwidth, clip = true]{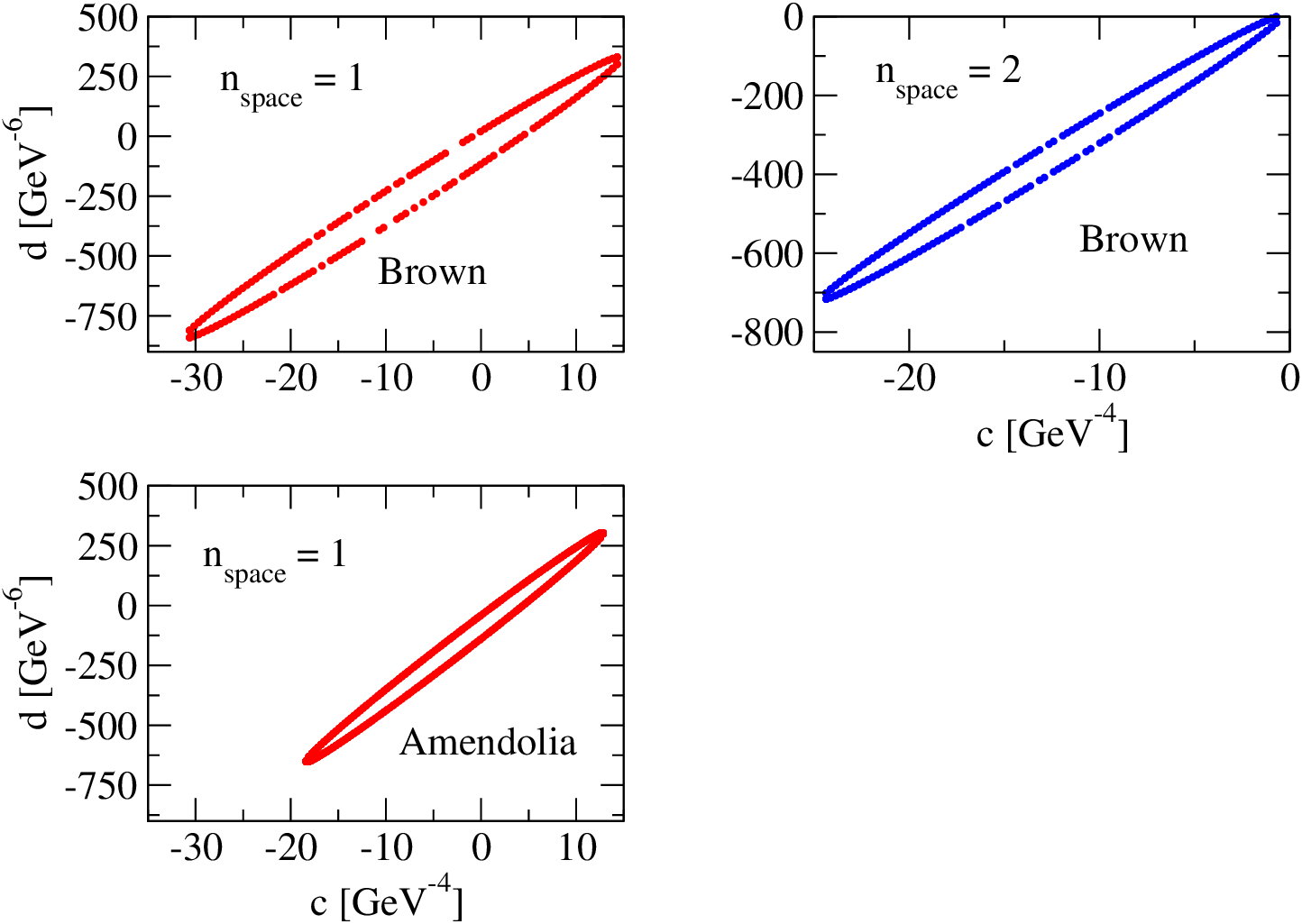}
\caption{Constraints on $c$ and $d$ as the number of space-like constraints are increased. Here the space-like data is taken from refs.~\cite{Brown},~\cite{Amendolia}.}
\label{cd_sep.fig2}
\end{center}
\end{figure*}

Including constraints from $a_3$ results in a relationship between the expansion coefficients $c$ and $d$ which we shall now explore. Fig.~(\ref{cd_sep.fig0}), shows the allowed region in the $c-d$ plane using only the normalization condition on $F_\pi(t)$ and the value of $r_\pi$. Analogous to the study of the bounds on $c$, we can include space-like constraints to impose stringent bounds on the expansion coefficients $c$ and $d$, that are reflected by smaller allowed regions in the $c-d$ plane.
Fig.~(\ref{cd_space1.fig2}) shows the variation in the bounds on $c$ and $d$ for one space-like constraint, where data from different $|t|$ region is used. In the figure, $h(x_1)$ is the value of $h(x)$ as defined in eqn.~(\ref{hz.eqn1}) corresponding to the smallest value of $t$ in each data set (refer tables~\ref{table_bebek} -~\ref{table_aman}). Similarly $h(x_2)$ is the constraint at a higher $|t|$ value taken in the ascending order of magnitude. It may be readily observed that the most
stringent bounds are obtained when data from smallest values of
$|t|$ are used, and they improve as the number
of constraints are increased, as noted earlier. 

\begin{figure*}
\begin{center}
\includegraphics*[angle = 0, width = 0.84\textwidth, clip = true]{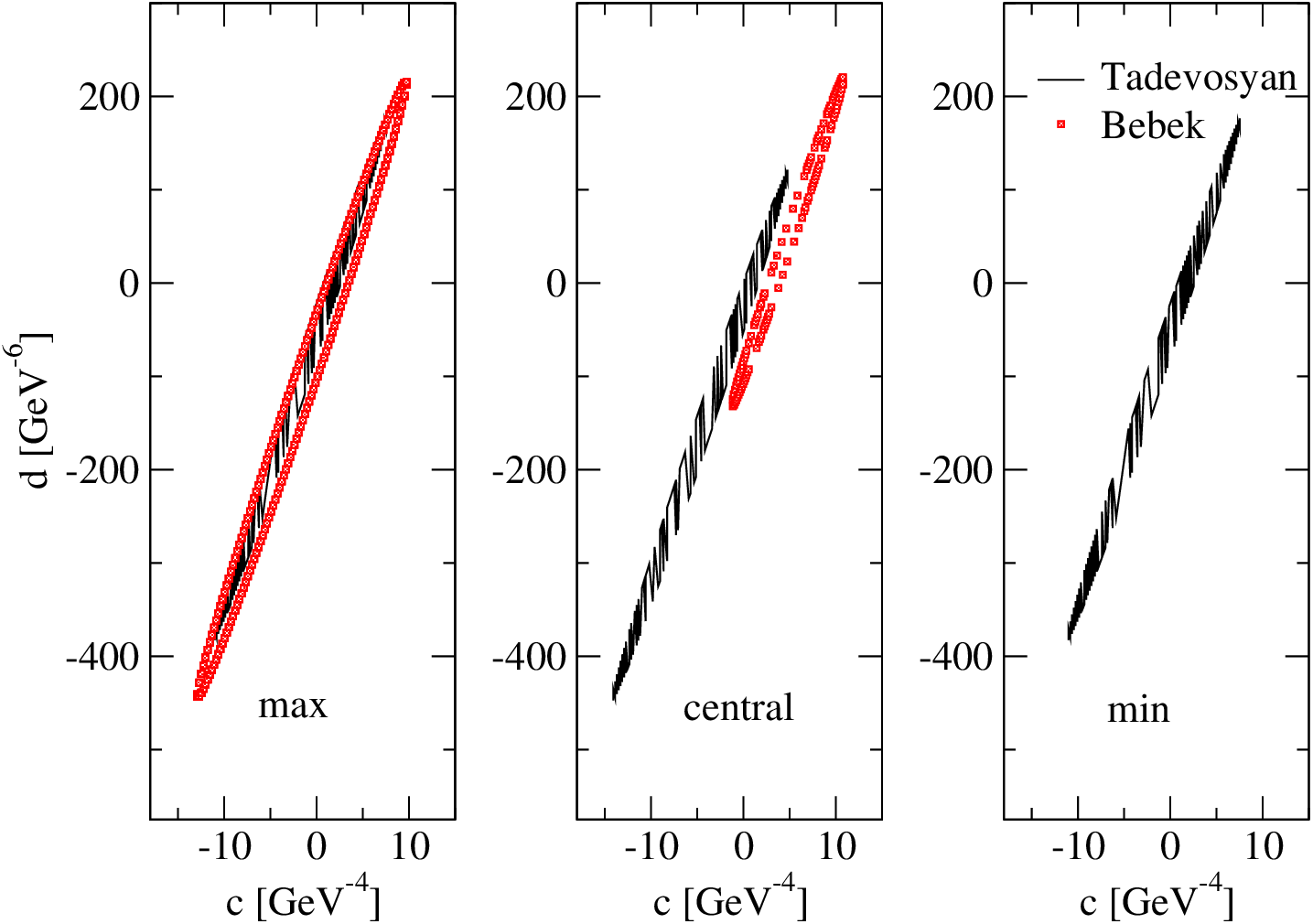}
\caption{Variations in the bounds on $c$ and $d$ as data from~\cite{Bebek2} and~\cite{nucl-ex/0607007} are varied within their error bounds for three space-like constraints. The shaded region represents the bounds from~\cite{nucl-ex/0607007} and the symbol represents those from~\cite{Bebek2}.}
\label{best.fig}
\end{center}
\end{figure*}

Fig.~(\ref{cd_sep.fig1}) (as well as fig.~(\ref{cd_sep.fig2})) shows that increasing the number of space-like constraints improves the bounds on the expansion coefficients $c$ and $d$. We observe that the allowed range for $c$ corresponds to that in fig.~\ref{cd_space1.fig1}. Using the central values tabulated in tables~\ref{table_bebek}-~\ref{table_aman}, we see that we are able to use up to three space-like constraint with the data from Bebek et al.~\cite{Bebek2} and the recent data from Tadevosyan et al.~\cite{nucl-ex/0607007}, while we are able to incorporate only two space-like constraints with the data from Brown et al.~\cite{Brown} and one with the data from Amendolia et al.~\cite{Amendolia}. The best estimate for the bounds on the Taylor coefficient $c$ and $d$ is obtained for the data from Bebek et al.~\cite{Bebek2}. But varying the Bebek data within the error bounds, we see that there is overlap with the bounds obtained with the data from Tadevosyan et al.~\cite{nucl-ex/0607007}, as seen in fig.~(\ref{best.fig}). In fig.~(\ref{best.fig}),  the label ``max'' refers to the data taken from the tables~\ref{table_bebek} and~\ref{table_jlab} with the corresponding error bounds added, i.e. at the upper limit of the error bound, central refers to the central values and ``min'' refers to central value minus the error bound, i.e., the lower limit of the error bound. We see that an overlap between the bounds obtained from the different data set occurs when the data are close to the upper error bound. In fact at the upper error bound, the data from Tadevosyan et al. gives better bounds on the expansion coefficients compared to the data from Bebek et al. At the central value of $F_\pi(t)$, the data from Bebek et al. does better while at the lower end, the data from Bebek et al., no longer yields stable result, while the data from Tadevosyan et al. still gives reasonable bounds. We find that despite small differences in the allowed regions, the fact 
that the allowed regions are essentially the same offers an important 
consistency check on the form factor determinations by each of the 
experiments.

As seen from the our results so far, the bounds on the Taylor coefficients vary due to the errors in the data presented in the tables~\ref{table_bebek} -~\ref{table_aman}; hence it is important to examine the sensitivity of our results to the inputs. We have made a systematic study of the sensitivity by varying, (a) 
the input value of $r_\pi$, 
(b) the input value of $I$,  and (c)
varying the experimental determinations of $F_\pi(t)$ within their
quoted errors, and in each instance keeping all
other inputs fixed. 

\bfig
\begin{center}
\includegraphics*[angle = 0, width = 0.44\textwidth, clip = true]{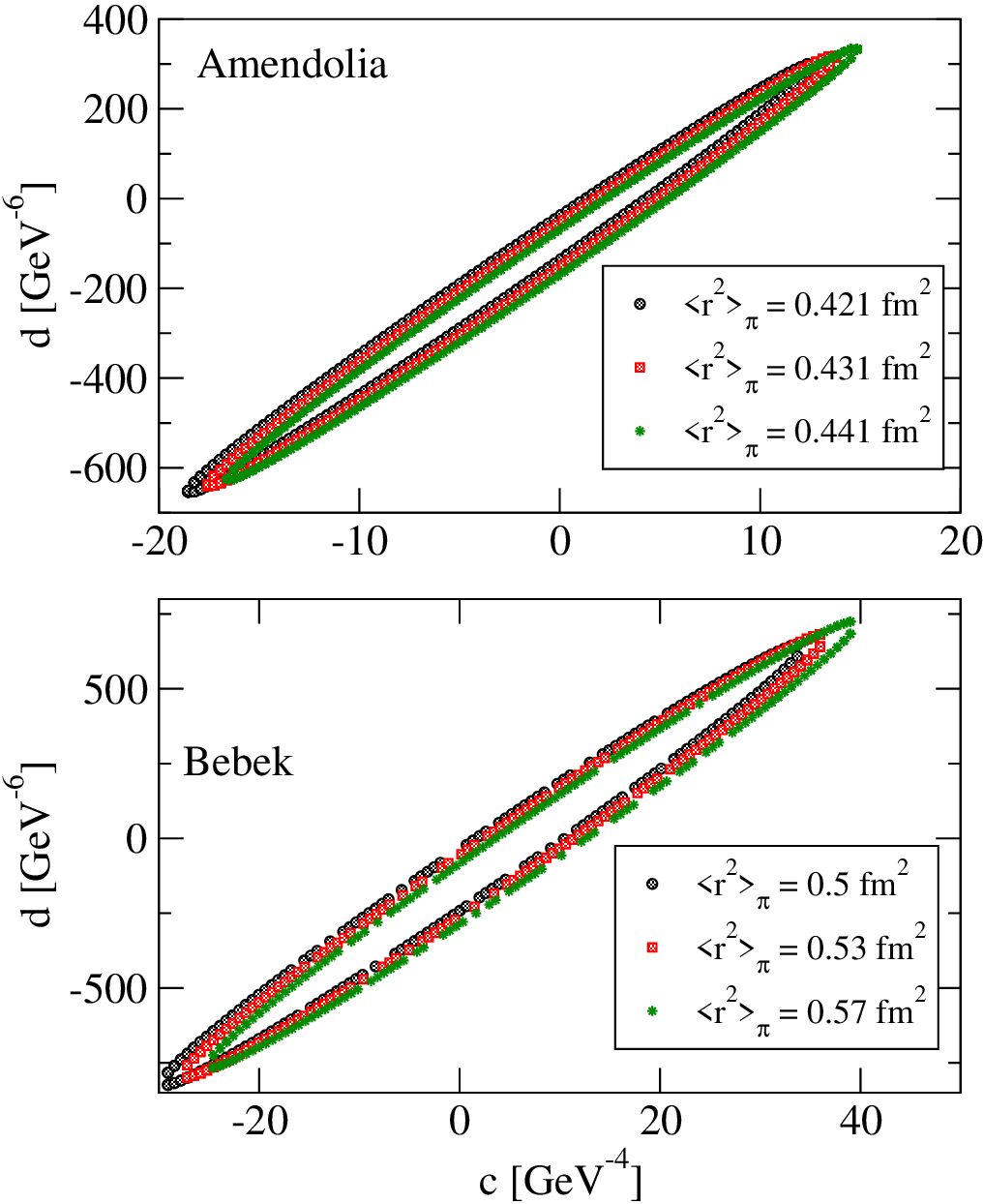}
\caption{Variations in the bounds on $c$ and $d$ when the value of $r_\pi$ is changed. This has been carried out for one space-like constraint for the data set by Amandolia et.al~\cite{Amendolia}.}
\label{aman_r_pi.fig}
\end{center}
\efig

\bfig
\begin{center}
\includegraphics*[angle = 0, width = 0.44\textwidth, clip = true]{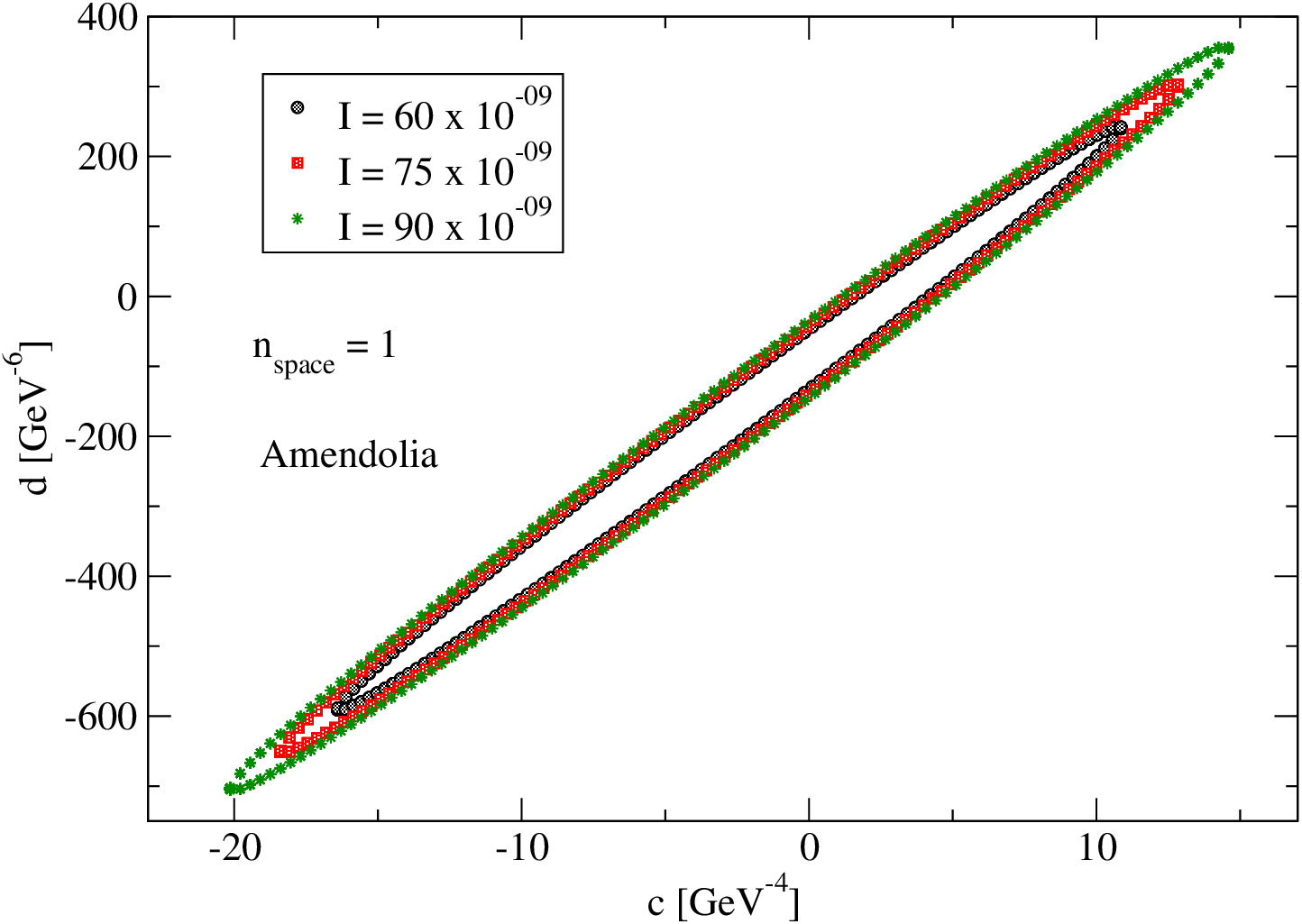}
\caption{Variations in the bounds on $c$ and $d$ when the value of $I$ is changed. This has been carried out for one space-like constraint for the data set by Amandolia et.al~\cite{Amendolia}.}
\label{aman_I.fig}
\end{center}
\efig

%It is important to examine the sensitivity of our results to the inputs.
Variations in the bounds on $c$ and $d$ for one space-like constraint from~\cite{Bebek2},~\cite{Amendolia}, when the value of $r_\pi$ is varied within the error bounds quoted therein, is shown in fig.~(\ref{aman_r_pi.fig}). We see that the bounds do not vary much as $r_\pi$ is varied within the allowed errors of a few percent.
Fig.~(\ref{aman_I.fig}) shows the variations in the bounds of $c$ and $d$ as the value of $I$ is changed. Lower $I$ values results in a subspace of the region allowed for the coefficients at higher $I$ values. We expect this behavior as our method gives only an allowed region for the coefficients. This explicitly demonstrates that as long as the true pionic contribution is lower than the values we have used, the allowed region lies within the isolated ellipse.

\bfig
\begin{center}
\includegraphics*[angle = 0, width = 0.44\textwidth, clip = true]{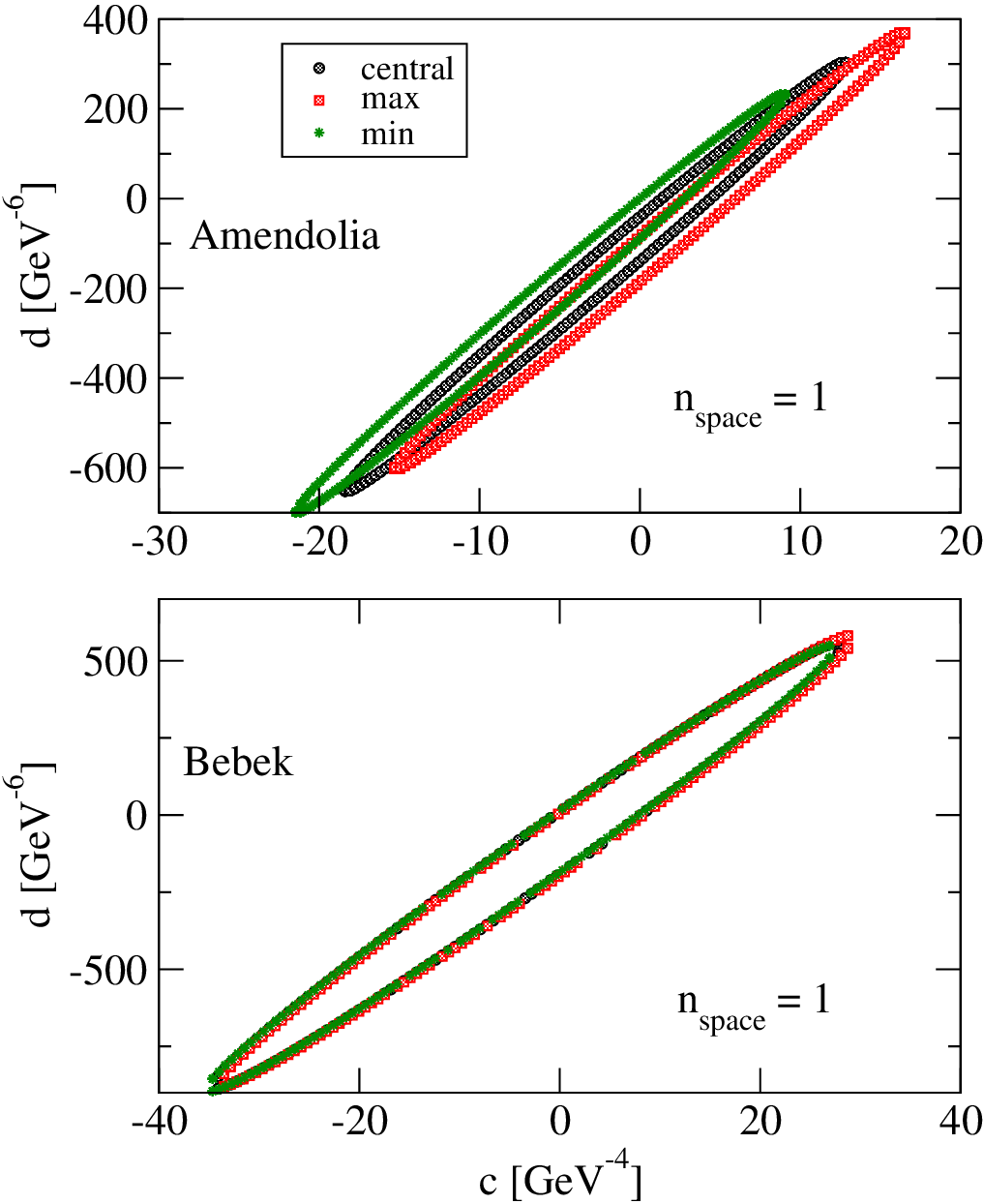}
\caption{Variations in the bounds on $c$ and $d$ when the data for $F_\pi(t)$ from tables~\ref{table_bebek} and~\ref{table_aman} are varied within their error bounds. Here ``central'' refers to the value without the errors and ``min'' to lowest value and ``max'' to the highest value within the error bounds. Note that the bounds are more sensitive to the errors in the data from the small $|t|$ region.}
\label{cd_err_bound.fig}
\end{center}
\efig

Fig.~(\ref{cd_err_bound.fig}) shows the variations bounds on $c$ and $d$ for the data from Amendolia et.al~\cite{Amendolia} and Bebek et.al~\cite{Bebek2} for one space-like constraint (smallest magnitude of $|t|$ in the tables~\ref{table_bebek} and~\ref{table_aman}). The data are varied within their experimental bounds and the results are depicted in fig.~(\ref{cd_err_bound.fig}). We see that the bounds on the expansion coefficients are most sensitive to errors in the data from lower $|t|$. Having said that, it is our view that one may read-off reliable ranges for both $c$ and $d$ in light of this sensitivity analysis. However an analogous sensitivity test that we have carried out for two space-like constraints for the data from Amendolia et al.~\cite{Amendolia} leads to a complete loss of coherence. This is an unavoidable numerical difficulty in the determinant that arises due to the fact that the entries are small and closely spaced. Therefore we can conclude that for low $|t|$ information we are unable to obtain reliable results for more than one constraint. 

\bfig
\begin{center}
\includegraphics*[angle = 0, width = 0.44\textwidth, clip = true]{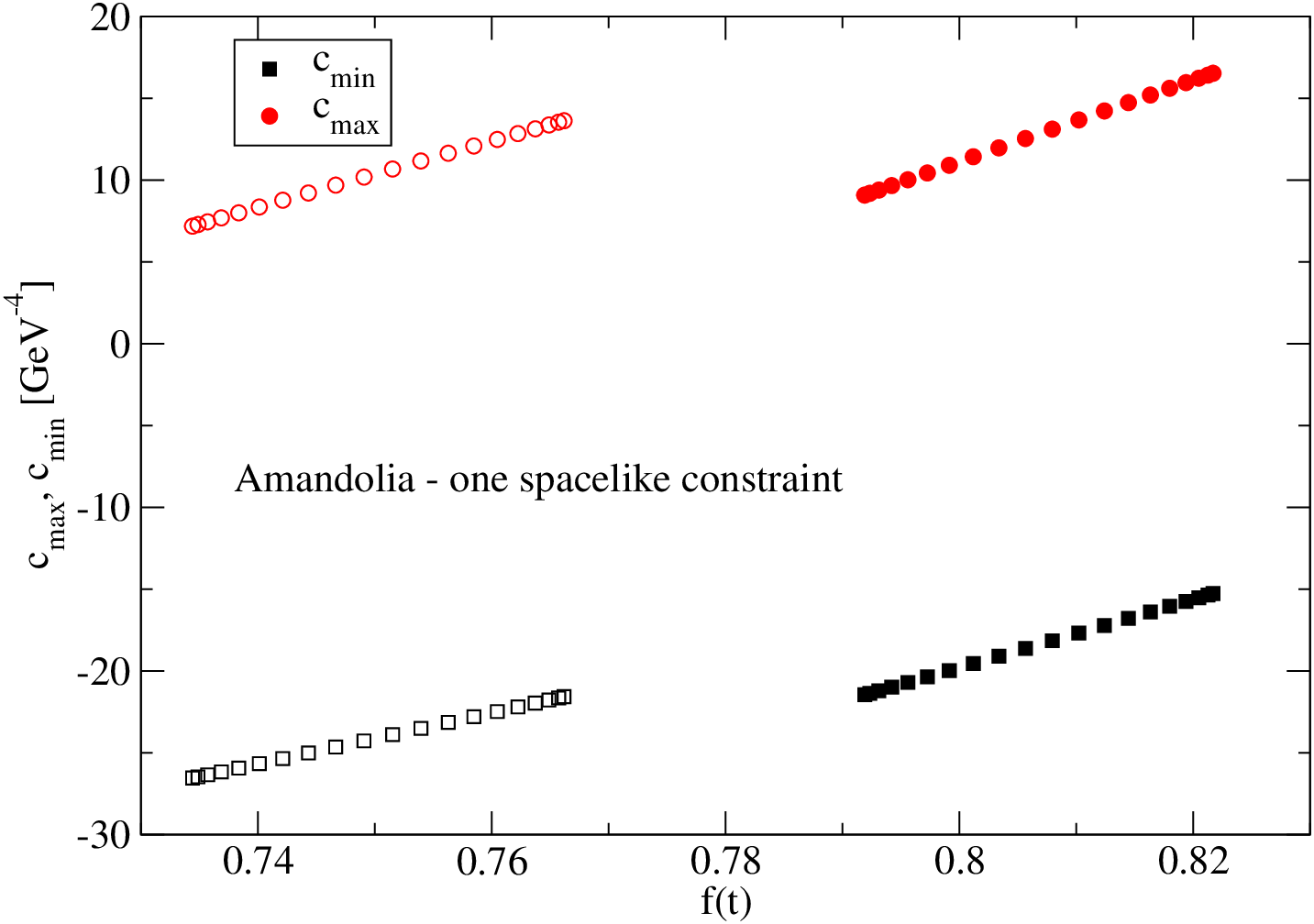}
\caption{Variations in the bounds on $c$ when the data for $F_\pi(t)$ from table~\ref{table_aman} are varied within their error bounds. The filled circles and squares refer to the maximum and minimum of the bounds on $c$ obtained from the first entry in table~\ref{table_aman} and the open ones correspond to the second entry in the table.}
\label{aman_ft.fig}
\vspace*{0.2cm}
\includegraphics*[angle = 0, width = 0.44\textwidth, clip = true]{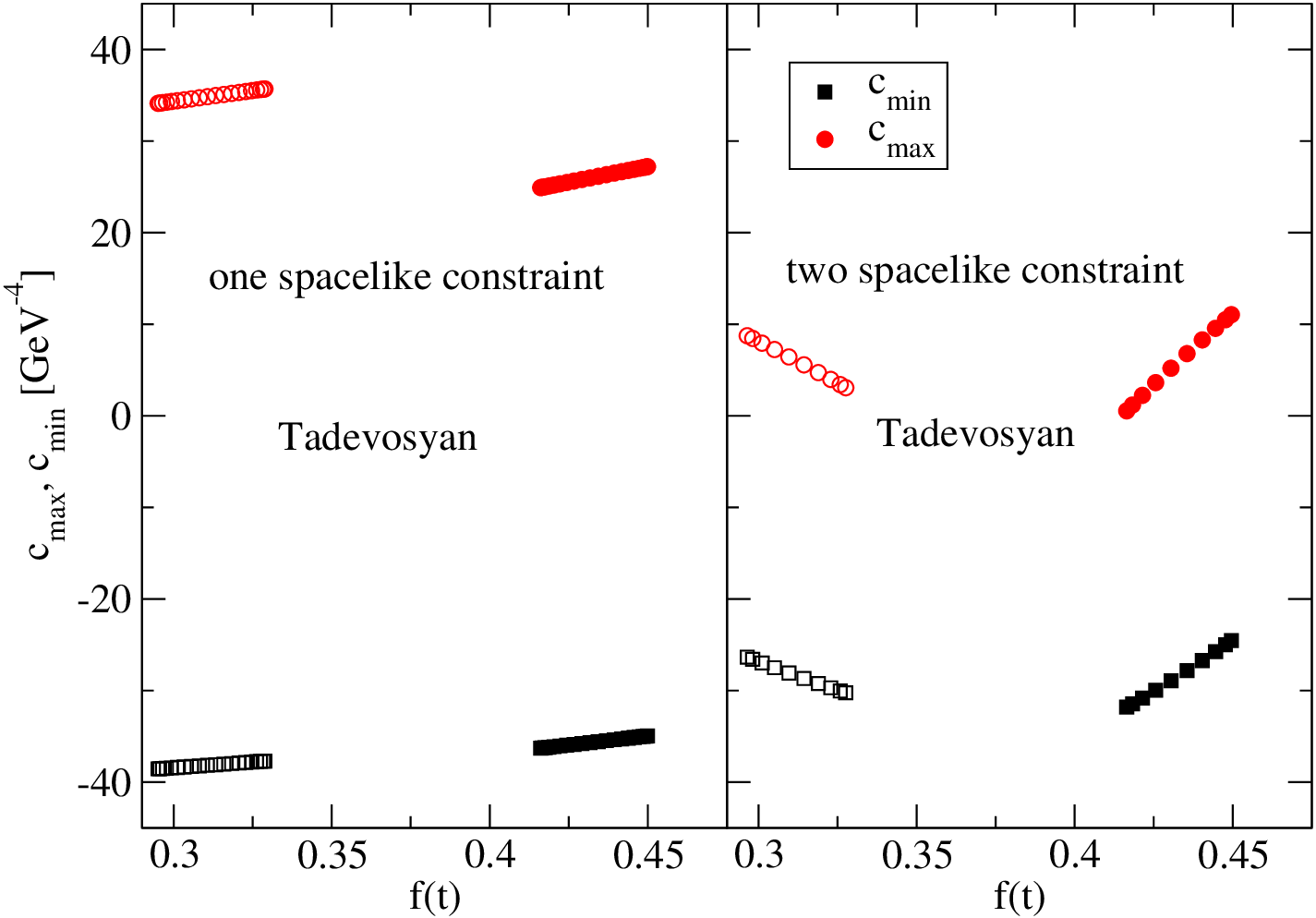}
\caption{Variations in the bounds on $c$ when the data for $F_\pi(t)$ from table~\ref{table_jlab} are varied within their error bounds. The filled circles and squares refer to the maximum and minimum of the bounds on $c$ obtained from the first entry in table~\ref{table_jlab} and the open ones correspond to the second entry in the table. The left panel shows the variation on one space-like constraint while the right panel shows variations on two space-like constraint. For two space-like constraints, we fix one constraint at the central value and vary the other over the quoted error bounds~\cite{Bebek2} and plot the variations in the bounds on $c$ as a function of the constraint that is varied.}
\label{JLAB_ft.fig}
\end{center}
\efig

We explore this sensitivity of the bounds to the data taken from the low $|t|$ region using one space-like constraint from~\cite{Amendolia} and varying this constraint over the given error bounds (Table~\ref{table_aman}) and study the corresponding variations in the bounds on $c$ alone. This is seen in fig.~(\ref{aman_ft.fig}) where the circles represent the maximum and the squares represent the minimum of the bounds. The filled symbols stand for the bounds obtained from the data at lower value of $|t|$ and the open symbols represent those from the data at higher value of $|t|$. Fig.~(\ref{JLAB_ft.fig}) shows similar variations in the bounds on $c$ when one as well as two space-like constraints (ref. table~\ref{table_jlab}) for the data taken from~\cite{nucl-ex/0607007} are varied over the error bounds. Note that the data from~\cite{nucl-ex/0607007} lie at higher $|t|$ value compared to the data from~\cite{Amendolia}. When two space-like constraints are used, we fix one constraint at the central value and vary the other over the error bounds quoted in~\cite{nucl-ex/0607007} and plot the maximum and minimum of the bounds as a function of the constraint that is varied. From figs.~(\ref{aman_ft.fig}) and~(\ref{JLAB_ft.fig}), we see that data from lower $|t|$ value in each set constraints the coefficients better for one space-like constraint. Also the sensitivity to the error bounds is greater for a set at lower $|t|$ compared to the set at higher value of $|t|$. We also note that two space-like constraints are more sensitive to the variations in $F_\pi(t)$ compared to one space-like constraint, as can be seen in fig.~\ref{JLAB_ft.fig}. 

\bfig
\begin{center}
\includegraphics*[angle = 0, width = 0.44\textwidth, clip = true]{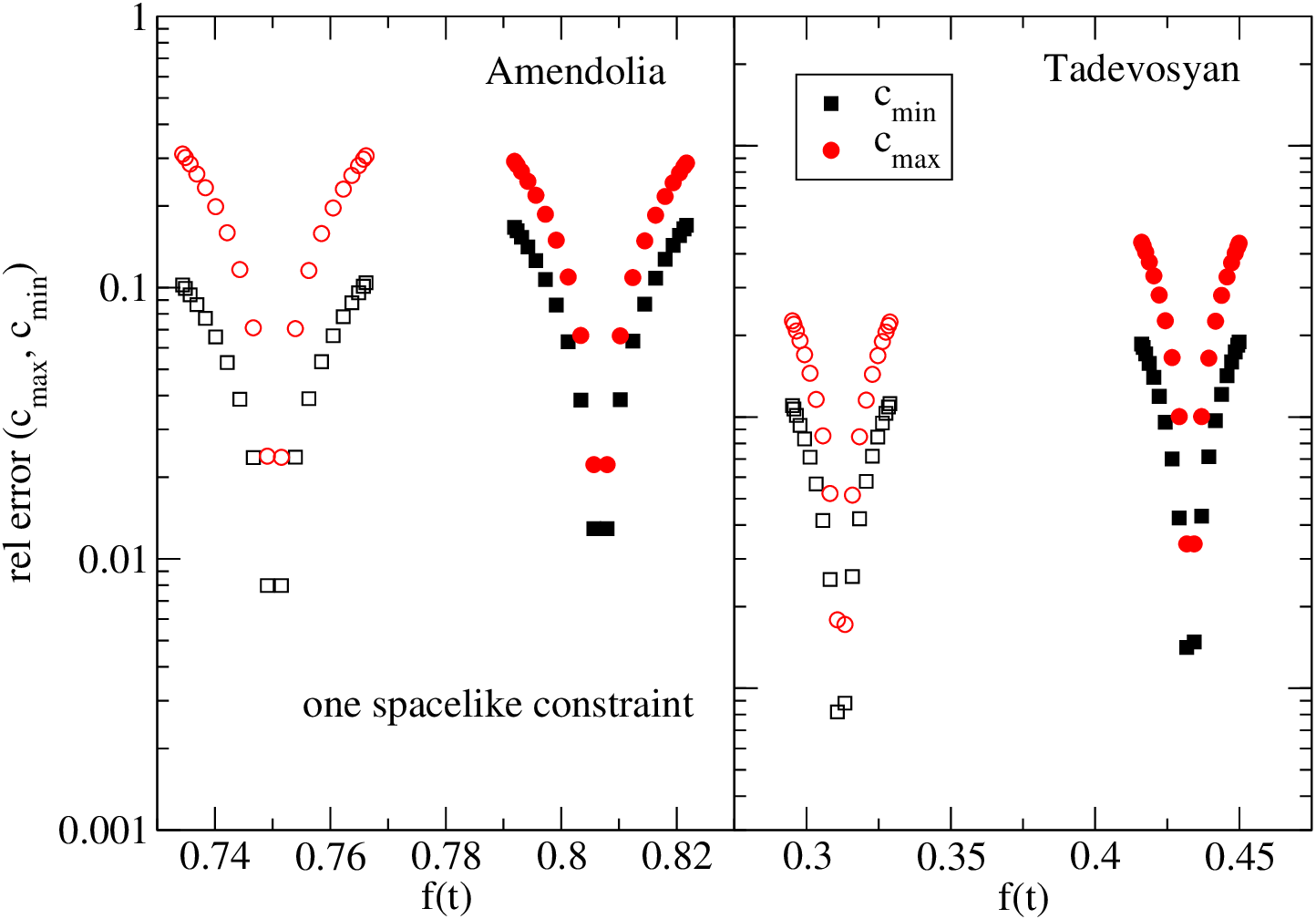}
\caption{Relative errors in the bounds on $c$ with respect to the central value, for one space-like constraint as the constraint is varied over the error bounds for the data sets~\cite{Amendolia} and~\cite{nucl-ex/0607007}.}
\label{rel_err.fig}
\end{center}
\efig

Relative errors in the bounds on $c$ as one space-like constraint is varied over the quoted error bounds compared to the bounds obtained from the central value is shown in fig.~(\ref{rel_err.fig}) for the data sets from Amendolia et al. and Tadevosyan et al.. We see that relative errors are smaller for the data from Tadevosyan et al., once again emphasizing our earlier observation that the sensitivity of the Taylor coefficients to the errors in the determination of $F_\pi(t)$ is greater for the data from the low $|t|$ region compared to those from the high $|t|$ region.  

%%%%%%%%%%%%%%%%%%%%%%%%%%%%%%%%%%%%%%%%%%%%%%%%%%%%%%%%%%%%%%%%%%%%%%%%%%%%%%%%%%%%%%%%%%
\section{Discussion and Conclusions}
\label{conclusions}
In this work we have considered the constraints on the pion 
electromagnetic form factor coming from present day estimates of the pionic 
contributions to the muon anomaly, and from space-like data which is 
available over a fairly extended kinematic regime. We have adopted the 
framework of Raina and Singh and have extended it to obtain constraints on 
$c$ and $d$, the expansion coefficients sub-leading to the charge radius of the 
pion.  We have used data available in the 1970's (also used in ref.~\cite{RainaSingh}) and more recent data which spans 
significantly lower values of $|t|$, as well as very recent data coming 
from Tadevosyan et al. which are at higher $|t|$ values, which lie in the range of Bebek et al.  

\bfig
\begin{center}
\includegraphics*[angle = 0, width = 0.44\textwidth, clip = true]{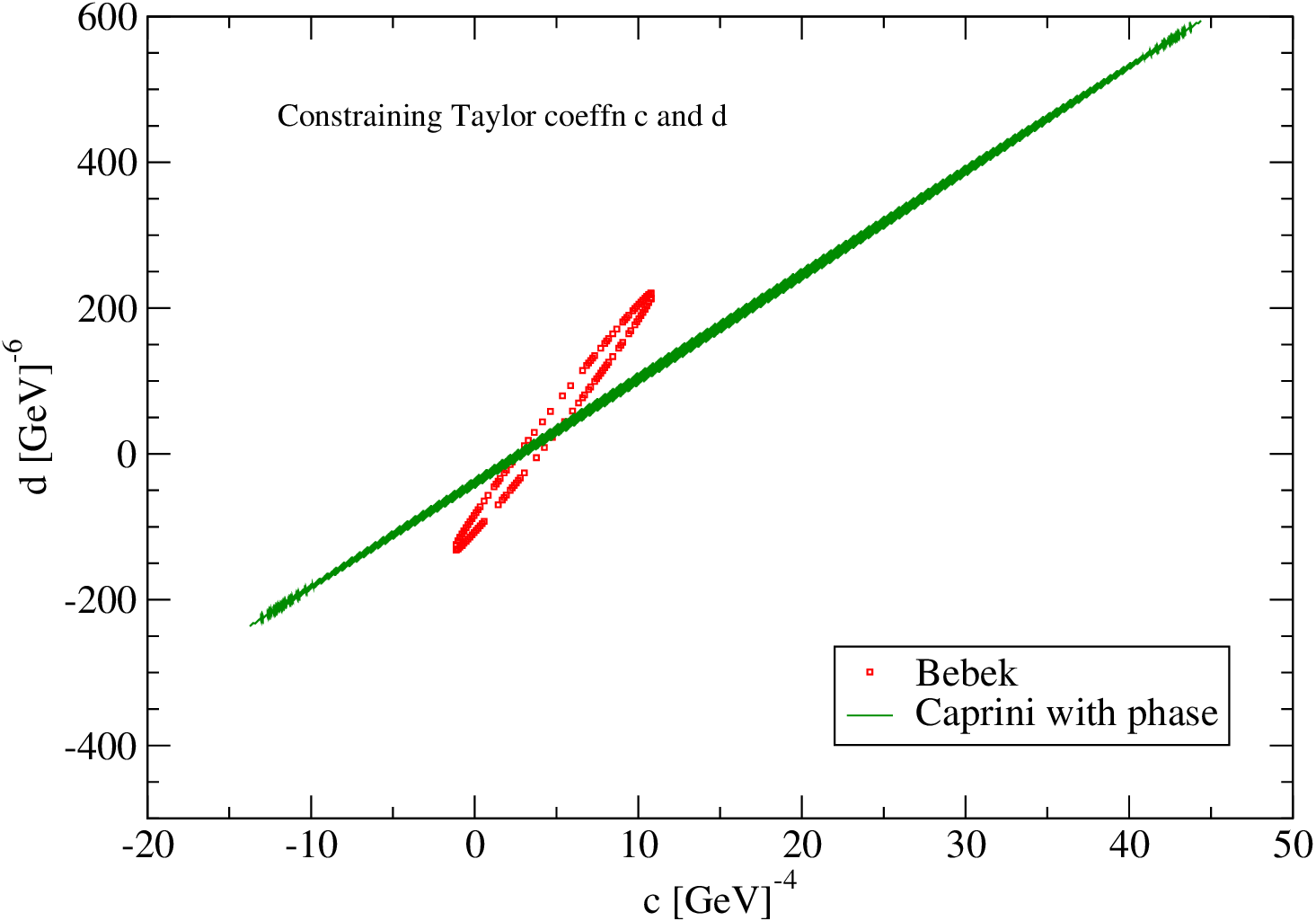}
\caption{Comparing our best ellipse obtained for three space-like constraints using data from Bebek et al.~\cite{Bebek2} (symbol) with the result obtained using the phase of the form factor in the timelike region (see fig.(2) in~\cite{Caprini}).}
\label{cap_RS.fig}
\end{center}
\efig

As mentioned in the Introduction, Caprini~\cite{Caprini} has used timelike data to obtain constraints on the Taylor coefficients. By systematically considering the
inclusion of only the phase, the ellipse with no constraints was found to shrink considerably using as an observable the QCD polarization function (see figs.(1) and (2) in~\cite{Caprini}).  This analysis 
employs an optimal technique, resulting in an integral equation
of the Fredholm type that was solved numerically.
The ellipse was found to shrink even further when both the phase
as well as the modulus information were used even though the technique
was non-optimal (see fig.(3) in~\cite{Caprini}). Questions remain about its validity
due to certain mathematical difficulties, as discussed by
Caprini.

The best estimates for the bounds on $c$ and $d$ is obtained using the data from Bebek et al.~\cite{Bebek2} for three space-like constraints. In fig.~(\ref{cap_RS.fig}), we compare our best result with those obtained by Caprini using only the phase from the timelike region. Using space-like data we get the following range for the expansion coefficients: $-1$ $\gev^{-4}$ $\lesssim c \lesssim$ $11 \gev^{-4}$ and $-132$ $\gev^{-6}$ $\lesssim d \lesssim$ $220$ $\gev^{-6}$. On the other hand Caprini has a range of [$-14$ $\gev^{-4}$, $44$ $\gev^{-4}$] and [$-236$ $\gev^{-6}$, $594$ $\gev^{-6}$] for $c$ and $d$ respectively using only the phase of timelike data. It is interesting to note that the overlap region between the two determinations, ours using space-like data and the muon anomaly and Caprini's using the phase of timelike data and the QCD polarization observable, accommodates comfortably the value of $c$ from chiral perturbation theory. 
%Thus it is possible that the use of refined data in the timelike region could lead to stringent constraints. 
We note here that similar bounds obtained using the central values of the Tadevosyan data (ref. table~\ref{table_jlab}) using three space-like constraints does not accommodate the value of $c$ obtained from chiral perturbation theory, but varying the data at the upper end of the error bound, as in fig.~\ref{best.fig}, the allowed region does include this value.

Our conclusions are that the constraints are weaker than those obtained by Caprini 
using both phase and modulus of data from the timelike region (see fig.(3) in~\cite{Caprini}), while they are
more stringent than those obtained with 
only the phase of the timelike data.  Here we note again, that the
latter of these treatments is rigorous and the results may be taken
as reliable.  On the other hand, the results obtained from using
both the modulus and phase while appearing more stringent are on
a less rigorous footing, due to inherent mathematical difficulties
as noted in ref.~\cite{Caprini}. 

%The simplicity and transparency of our 
%results are worth emphasizing.

We have also carried out a sensitivity analysis by varying the 
estimate for the pionic contribution over a significant range, varying the 
charge radius over its presently known errors, and also by varying the 
experimental data that we use over its errors, for the values of $|t|$ 
that we have chosen. Our conclusions are that in the small $|t|$ region, 
the system is very sensitive and as a result, we have been unable to implement 
anything more than one space-like constraint.  For larger values, we are 
able to include two and even three constraints.  

Of related interest, is the extreme 
sensitivity of the system when bounds are being derived on the
muon anomaly from space-like data, as recognized earlier
by Pantea and Raszillier~\cite{Pantea:1977ge}.
The issue was of some significance 
because several space-like constraints were being used to obtain these 
bounds.  Our circumstances are somewhat mitigated by the fact that we are 
using significantly smaller number of constraints, as the determinantal 
equations we are solving are already of rather large dimensions.  By 
studying the sensitivity of the bounds, we are confident that our results remain stable.

It is conceivable that as data improves one may obtain better constraints 
on $c$ and $d$, where it is numerically feasible, that can then be used as inputs for high precision, 
self-consistent form factor determinations. We also note at this point that our work takes into account only the information present in the space-like region. It would be worth-while to properly formulate the problem so that the information available in the timelike region could be incorporated together with a comprehensive error analysis. A theory of error functionals has been developed by Raina and Singh~\cite{RS_NPB} in the context of finding a lower bound for the muon anomaly, that
could in principle be extended for the problem at hand, which is beyond the 
scope of this work.  Therefore, future investigations could combine highly 
accurate phase and modulus timelike information coming from recent 
experiments and the techniques developed by Caprini for the phase problem,
together with a suitable extension of the error functional method, to produce 
stringent constraints on $c$ and $d$.  

%%%%%%%%%%%%%%%%%%%%%%%%%%%%%%%%%%%%%%%%%%%%%%%%%%%%%%%%%%%%%%%%%%%%%%%%%%%%%%%%%%%%%%%%%%
 
\begin{acknowledgement}
We are indebted to I.\ Caprini for careful reading and comments on the manuscript and several patient discussions.
We also thank G.\ Colangelo, H.\ Leutwyler, B.\ Moussallam and A. Upadhyay for discussions.
BA thanks the Department of Science and Technology,
Government of India, for support. BA also thanks V.\ Singh for drawing attention to refs.~\cite{SinghRaina},~\cite{RainaSingh} and~\cite{RS_NPB}.
\end{acknowledgement}

%\section*{References}

\end{document}